\documentclass[aps,twocolumn,epsf,eps,floats,pre,nofootinbib,showpacs]{revtex4-1}
\usepackage[usenames]{color} 
\usepackage{amssymb,amsfonts,amsmath}
\usepackage{acronym}
\usepackage{graphicx}

\def\TMC{T_{MC}} 
 
\def\xs{\xi_{\rm s}}

\begin{document}

\title{Dynamic relaxation of a liquid cavity under amorphous boundary
  conditions}

\author{Andrea Cavagna}
\affiliation{Istituto Sistemi Complessi (ISC), Consiglio Nazionale delle Ricerche (CNR), UOS Sapienza, Via
  dei Taurini 19, 00185 Roma, Italy} 
\affiliation{Dipartimento di Fisica, Universit\`a\ ``Sapienza'', P.le
  Aldo Moro 2, 00185 Roma, Italy}

\author{Tom\'as S.\ Grigera}
\email{tgrigera@inifta.unlp.edu.ar}
\affiliation{Instituto de Investigaciones
  Fisicoqu{\'\i}micas Te{\'o}ricas y Aplicadas (INIFTA) and
  Departamento de F{\'\i}sica, Facultad de Ciencias Exactas,
  Universidad Nacional de La Plata, c.c. 16, suc. 4, 1900 La Plata,
  Argentina}
\affiliation{CONICET La Plata, Consejo Nacional de
  Investigaciones Cient{\'\i}ficas y T{\'e}cnicas, Argentina}

\author{Paolo Verrocchio}
\email{paolo.verrocchio@unitn.it}
\affiliation{Dipartimento di Fisica, Universit{\`a} di Trento, via
  Sommarive 14, 38050 Povo, Trento, Italy}
\affiliation{Istituto Sistemi Complessi (ISC-CNR), UOS Sapienza, Via
  dei Taurini 19, 00185 Roma, Italy} 
\affiliation{Instituto de Biocomputaci\'on y F\'{\i}sica de Sistemas Complejos (BIFI), Spain}

%% ABSTRACT %%%%%%%%%%%%%%%%%%%%%%%%%%%%%%%%%%%%%%%%%%%%%%%%%%%%%%%%%%%%%

\begin{abstract}
  The growth of cooperatively rearranging regions was invoked long ago
  by Adam and Gibbs to explain the slowing down of glass-forming
  liquids. The lack of knowledge about the nature of the growing
  order, though, complicates the definition of an appropriate
  correlation function. One option is the point-to-set correlation
  function, which measures the spatial span of the influence of
  amorphous boundary conditions on a confined system.  By using a swap
  Montecarlo algorithm we measure the equilibration time of a liquid
  droplet bounded by amorphous boundary conditions in a model
  glass-former at low temperature, and we show that the cavity
  relaxation time increases with the size of the droplet, saturating
  to the bulk value when the droplet outgrows the point-to-set
  correlation length.  This fact supports the idea that the
  point-to-set correlation length is the natural size of the
  cooperatively rearranging regions.  On the other hand, the cavity
  relaxation time computed by a standard, nonswap dynamics, has the
  opposite behavior, showing a very steep increase when the cavity
  size is decreased.  We try to reconcile this difference by
  discussing the possible hybridization between MCT and activated
  processes, and by introducing a new kind of amorphous boundary
  conditions, inspired by the concept of frozen external state as an
  alternative to the commonly used frozen external configuration.
\end{abstract}


\maketitle

%%
%% Acronyms (needs to go after \maketitle, at least for the twocolumns option)
%%

\acrodef{IS}{inherent structure}
\acrodef{RFOT}{random first-order theory}
\acrodef{CRR}{cooperatively rearranging region}
\acrodef{MCT}{mode-coupling theory}
\acrodef{MC}{Monte Carlo}
\acrodef{KCM}{kinetically constrained models}
\acrodef{SPM}{square plaquette model}
\acrodef{TPM}{triangular plaquette model}
\acrodef{ABC}{amorphous boundary conditions}
\acrodef{PBC}{periodic boundary conditions}

%% INTRO  %%%%%%%%%%%%%%%%%%%%%%%%%%%%

\section{Introduction}

It is common wisdom that the spectacular slowing down of supercooled
liquids at low temperature is caused by the growth of a correlation
length of some sort. The underlying idea is that of cooperativity: at
lower temperatures, larger regions (termed cooperatively rearranging
regions) must move together in order to fully relax
\cite{glassthermo:adam65}.  Unfortunately, the standard tools used in
critical phenomena to detect a growing correlation length fail in
glass-forming liquids, as it is not at all clear {\sl a priori} what
the order parameter should be. No obvious domain or structure can be
observed in a low temperature liquid to distinguish it from a high
temperature one.  If order is growing in glass-formers, it must be
some sort of amorphous order, and the corresponding order parameter
must be nonstandard.

Recently, some progress has been achieved by using \ac{ABC}
\cite{mosaic:bouchaud04, self:prl07, self:nphys08}.  The idea goes as
follows \cite{mosaic:bouchaud04}. Consider a low-temperature
equilibrium configuration of a liquid and freeze all particles outside
a certain region. This region (or cavity) is then let free to evolve
and thermalize, subject to the pinning field produced by all the
frozen particles surrounding it.  Clearly, the smaller the region the
stronger the effect of the pinning field, hence keeping the region in
a very restricted portion of its own phase space. The idea, then, is
to check how large the region must be to emancipate from the boundary
conditions, {\sl i.e.\/} to regain ergodicity and thermalize into a
state different from the surrounding one. The advantage of this method
is that the system chooses its own definition of `order' by means of
the amorphous boundary conditions, and we do not need to have any {\it
  a priori} knowledge of the nature of such order.  Practically
speaking, the procedure amounts to measure, as a function of the size
$R$ of the region, the correlation between the original region's
configuration (that of the frozen surrounding) and that achieved after
the region has equilibrated subject to the \acl{ABC}. This quantity is
called point-to-set correlation function \cite{cs:mezard06,
  dynamics:montanari06}, $q(R)$, and it has shown an interesting
feature \cite{self:prl07,self:nphys08}: its decay length-scale,
$\xi_s$, increases on lowering $T$.  Regions smaller than $\xi_s$
\emph{cannot} relax completely, even given infinite time, due to the
presence of the pinning \ac{ABC}.

Here, in order to get some information about the dynamics of the
cooperatively rearranging regions, we study the dynamical behavior of
a cavity under ABC. Of course, we do expect that the equilibration
time of the cavity must be equal to its bulk value for large enough
values of $R$. What is not trivial is at what specific value of $R$
the saturation occurs and whether the saturation occurs from above or
from below, {\sl i.e.} whether the equilibration time decreases or
increases when the cavity gets larger.  As we shall see, we obtain
different results according to the specific dynamics we use: by means
of a swap Montecarlo dynamics, where particles of different species
can be exchanged in order to accelerate the dynamics, we find a clear
saturation from below of the equilibration time taking place at
$R\sim\xi_s$ (Section~\ref{sec:swap-dynam-conf}). This result seems to
support the idea that $\xi_s$ is indeed the cooperativity length scale
of the system
(Sections~\ref{sec:rfot-interpr-swap}--~\ref{sec:an-unexp-ineq}).  If,
on the other hand, we use a standard, nonswap algorithm, we find that
the dynamics slows down very steeply when the cavity size is decreased
(Section~\ref{sec:nonsw-dynam-conf}).

We will discuss the meaning of these results and how this difference
may be related to the actual relaxation mechanism active for the
cooperatively rearranging regions in the bulk. In particular, we will
show that the hybridization between the Mode Coupling Theory (MCT) and
the activated relaxation channels can give rise to a nonmonotonic
behavior of the relaxation time as a function of the cavity radius
(Section~\ref{sec:contr-betw-swap}). We will also put forward the
hypothesis that the standard amorphous boundary conditions, where the
external configuration is frozen, introduce an artificial slowing down
of the cavity dynamics, which can be overcome by switching to the more
physical \emph{frozen state} boundary conditions
(Sections~\ref{sec:fc-vs-vs}--~\ref{sec:cavity-dynamics-with}). After
briefly commenting on some relevant experiments
(Section~\ref{sec:some-exper-evid}) we summarize our conclusions in
Section~\ref{sec:conclusions}.

%%%%%%%%%%%%%%%%%%%%%%%%%%%%%%%%%%%%%%%%%%%%
\section{Model and observables}
%%%%%%%%%%%%%%%%%%%%%%%%%%%%%%%%%%%%%%%%%%%%

We perform \ac{MC} simulations of a 3-$d$ soft-spheres binary mixture
\cite{soft-spheres:bernu87} with parameters as in
Ref.~\cite{self:nphys08}. The mode-coupling temperature
\cite{rev:goetze92} for this system is $\TMC=0.226$
\cite{soft-spheres:roux89}.  Our largest system has $N=16384$
particles in a box of length $L=25.4$.  We run simulations at
$T=0.482,0.350,0.246,0.214,0.202$. We first equilibrate the whole
system with periodic boundary conditions to generate a set of
equilibrium configurations, and then run the amorphous boundary
simulations by picking an equilibrium configuration and artificially
freezing all particles but those occupying a spherical cavity of
radius $R=$ 1.06, 1.68, 1.92, 2.12, 2.28, 2.61, 2.87, 3.29, 3.62, 4.15,
4.57, 5.75, 7.2, 9.14, and 10.95. We use at least 16 samples for each
$T$ and $R$.

Our main physical observable is the overlap, which measures the
correlation between the running configuration and the reference one at
$t=0$. The cavity is partitioned in small cubic boxes and $n_i$ is the
number of particles in box $i$. The side $\ell$ of the cells is such
that $n_i=\{0,1\}$. We measure the overlap within a small cubic volume
$v$ located at the center of the sphere \cite{self:nphys08},
\begin{equation}
q(t;R)\equiv \frac{1}{\ell^3 \, N_i} \sum_{i \in
v} n_i(t) \, n_i(0)  \ ,
\end{equation}
where the sum runs over all boxes and $N_i$ is the number of boxes in
the central volume. To minimize statistical uncertainty without losing
the local nature we choose $N_i = v/\ell^3 = 125$. On average, the
overlap of two identical configurations is $1$, while for totally
uncorrelated configurations $q = q_0 = \ell^3 = 0.062876$.  The
asymptotic value of the overlap, $q(R)\equiv \langle
q(t\to\infty;R)\rangle$, averaged over many realizations of the
boundary conditions, is the point-to-set correlation function
\cite{mosaic:bouchaud04,
  dynamics:montanari06,self:prl07,self:nphys08,review:biroli09}.

In order to define a time-scale we measure the connected auto-correlation function of the
overlap fluctuations,
\begin{equation} 
C(t;R) =\frac{
\left\langle \left( q(t_0+t;R)- q(R) \right) \left( q(t_0;R)- q(R)
\right) \right\rangle  }
{
\left\langle \left( q(t_0;R)- q(R) \right)^2  \right\rangle
}
\ .
\label{connessa}
\end{equation}
To estimate the equilibration time $\tau(R)$ we use the method
discussed by A.~Sokal in \cite{review:sokal97}, based on the integral of the
correlation function. More specifically, the relaxation time $\tau$ is
found by solving the equation,
\begin{equation}
\tau = \int_0^{\alpha \tau} dt \ C(t;R)  \ ,
\end{equation}
where the optimal value of $\alpha$ has been found to be $20$. In this
way one is sure to sample the phenomenon on a time window that is self
consistently much larger than the relaxation time.

%%%%%%%%%%%%%%%%%%%%%%%%%%%%%%%%%%%%%%%%%%%%
\section{Swap dynamics in the confined cavity}
%%%%%%%%%%%%%%%%%%%%%%%%%%%%%%%%%%%%%%%%%%%%

\label{sec:swap-dynam-conf}

\begin{figure}
  \includegraphics[angle=-90,width=\columnwidth]{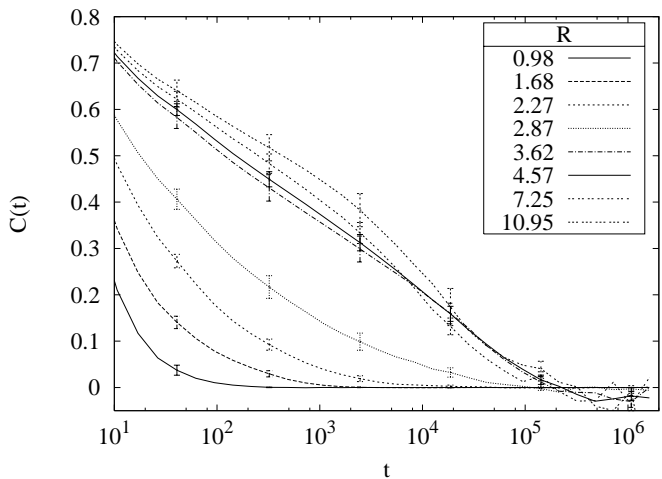}
  \caption{Autocorrelation function $C(t;R)$ for a few representative
    sizes $R$ at $T=0.202$. }
  \label{fig:corr-allT}
\end{figure}

We first focus on the results obtained with swap dynamics
\cite{self:pre01}. With a swap Monte Carlo dynamics we propose (with
probability 0.1) a move that swaps the position of two different
particles. Provided that the radii of the two different species are
not too different, so that the swap move is not always rejected, this
kind of move decreases significantly the time needed by a single
particle to break its cage. On the other hand, the swap dynamics has
less of an impact on collective rearrangements, and indeed the swap
relaxation time increases dramatically close to the glass transition,
as the nonswap time.

Fig.~\ref{fig:corr-allT} shows the swap auto-correlation function
$C(t;R)$ at various values of $R$ for our lowest temperature
$T=0.202$.  We stress that for those values of $R$ such that the order
parameter $q(R)\neq 0$, ergodicity is broken~\cite{self:nphys08}. In
this case the {\it connected} correlation function~\eqref{connessa}
describes the equilibrium dynamics \emph{within a restricted region}
of the cavity's phase space.

\begin{figure}
  \includegraphics[angle=-90,width=\columnwidth]{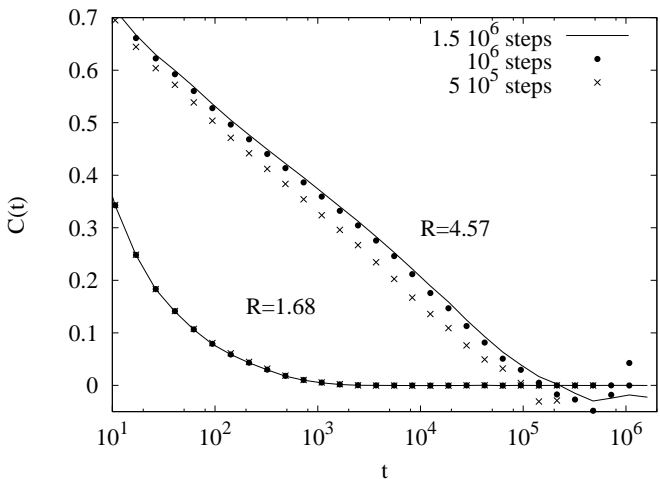}
  \caption{Test of thermalization of the autocorrelation function. We
    have computed $C(t,R)$ using increasingly longer time windows, to
    be sure that the autocorrelation function was saturated. Neither
    for large or small $R$ is there evidence of any residual
    dependence on the time window. $T=0.202$.}
  \label{fig:fss}
\end{figure}

Before estimating the relaxation time it is very important to be sure
that the autocorrelation function does not depend on the size of the
time window $\Delta t$ used to measure it. To this aim in
Fig.~\ref{fig:fss} we show the autocorrelation function at our lowest
temperature and at different values of the time window $\Delta t$, at
two values of $R$: there is no significant dependence of $C(t;R)$ on
$\Delta t$.

\begin{figure}
\includegraphics[angle=-90,width=\columnwidth]{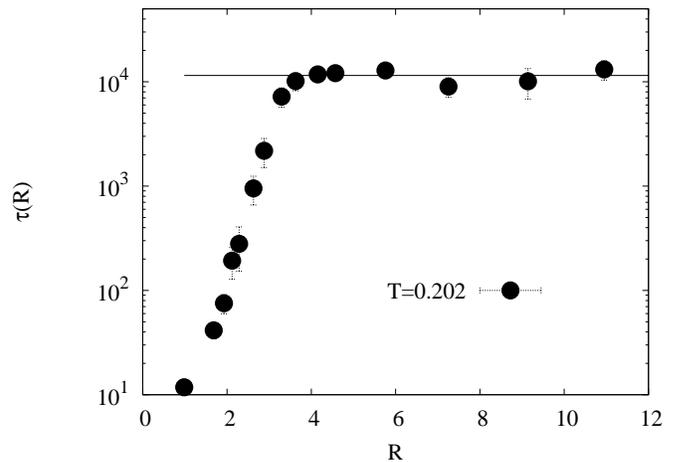}
\caption{Cavity relaxation time vs.\ $R$ for $T=0.202$. The kink
  between the growth and the saturated regime occurs close to $\xi_s =
  3.82\pm 0.46$, which is the point-to-set lengthscale for $T=0.202$
  \cite{self:nphys08}. Also shown is the bulk (periodic boundary
  conditions) equilibration time (full line).}
\label{fig:tau149}
\end{figure}

In Fig.~\ref{fig:tau149} we report the swap equilibration time
$\tau(R)$ for our lowest temperature, $T=0.202$. Three features of
this curve stand out: i) the swap equilibration time saturates for
large $R$ to a value independent of the cavity size; ii) the swap
equilibration time grows with $R$, so that saturation occurs from
below; iii) growth and saturation are separated by a rather sharp kink
at a well-defined value of $R$. The first fact is obvious: the effect
of the boundary conditions is expected to fade away for large $R$, so
that $\tau(R)$ must eventually reach its bulk value, which is exactly
what happens.  The remarkable point is that $\tau(R)$ reaches its bulk
value for $R \sim \xi_s$, where $\xi_s$ is the point-to-set
correlation length measured in Ref.~\cite{self:nphys08}.

This result can immediately be interpreted in terms of cooperativity:
For $R<\xi_s$ the whole region is correlated, because the effect of
the amorphous border breaks the ergodicity. For $R>\xi_s$, the effect
of the border fades away and the region is able to decorrelate by
breaking up into smaller correlated sub-parts: in this regime
relaxation factorizes. Hence, it seems that the point-to-set
correlation length $\xi_s$ does indeed play a role in the cooperative
dynamics of the system. In the next Sections we will address this
point more precisely.

%%%%%%%%%%%%%%%%%%%%%%%%%%%%%%%%%%%%%%%%%%%%
\section{RFOT interpretation of the swap equilibration time}
%%%%%%%%%%%%%%%%%%%%%%%%%%%%%%%%%%%%%%%%%%%%

\label{sec:rfot-interpr-swap}

According to the \ac{RFOT} of the glass transition, whether or not a
region of radius $R$ relaxes depends on the balance between the
surface tension $Y$ that develops when that region actually rearranges
and the configurational entropy $\Sigma$ unleashed by the
rearrangement: if $Y>T\Sigma R^{d-\theta}$ ($d$ is the space
dimension, $\theta$ is the surface tension ---or stiffness---
exponent) the surface cost is larger than the entropic gain and the
region does not rearrange.  On the other hand, if $Y<T\Sigma
R^{d-\theta}$ the entropic gain outweighs the surface energy cost and
the region has a thermodynamic advantage to rearrange. The rearranging
size where entropy and surface tension balance, $\xi_s=
\left(Y/T\Sigma\right)^{1/(d-\theta)}$, is the static correlation
length of \ac{RFOT}.

Therefore, within \ac{RFOT} a cavity with amorphous boundary
conditions of radius $R<\xi_s$ has broken ergodicity, and can only
explore the state imposed by the boundary conditions
\cite{mosaic:bouchaud04}. In this regime the equilibration time is the
time needed to explore that one state, which is roughly equal to the
$\beta$-relaxation time, $\tau(R)\sim \tau_\beta$ \footnote{We neglect
  in this analysis a possible dependence of $\tau_\beta$ on $R$ due to
  the extended nature of the excitations related to $\beta$-relaxation
  \cite{mosaic:stevenson10}.}.  For $R>\xi_s$, instead, rearrangement
occurs and ergodicity of the cavity is restored. In this regime the
region is larger than the minimal rearranging size, so that relaxation
factorizes: different subregions of size $\xi_s$ will rearrange
independently from each other, and the equilibration time will be
equal to its bulk value, i.e. $\tau(R)\sim \tau_0 \exp(\xi_s^\psi/T)$,
where $\tau_0$ is an Arrhenius prefactor and $\psi$ is the exponent
regulating the barrier growth.

Hence, within the sharp \ac{RFOT} description, where the surface
tension has just one value, $Y$, it is predicted a step-like jump of
$\tau(R)$ at $R=\xi_s$, from $\tau_\beta$ up to $\tau_0
\exp(\xi_s^\psi/T)$,
\begin{equation}
\tau(R) \sim
  \begin{cases}
    \tau_\beta & R< \xi_s \\
     \tau_0\, e^{\xi_s^\psi/T}
  & R>\xi_s   \ .
  \end{cases}
\label{assuppa}
\end{equation}
\noindent
Such stepwise behavior is not what we observed in
Fig.~\ref{fig:tau149}. In order to reconcile data and theory, we note
that for the typical temperatures and sizes studied in simulations
 surface tension fluctuations are
relevant  \cite{self:nphys08}. 
If the surface tension fluctuates \footnote{ In fact, both surface tension and
  configurational entropy will fluctuate \cite{mosaic:dzero05}.  At
  the practical level, though, disentangling the two effects is hard,
  and given that large surface tension fluctuations have been reported
  \cite{self:jstatmech09,self:jcp09a}, a generalized version of
  \ac{RFOT} that incorporates only surface tension fluctuations seems
  reasonable \cite{self:nphys08}.}  
(i.e.\ different \acp{ABC} give different $Y$), local excitations can
have different sizes and therefore different relaxation times. When we
measure these quantities by averaging over many different sets of ABC
we smooth out the sharp behavior of \eqref{assuppa}.

\begin{figure}
\includegraphics[angle=0,width=\columnwidth,clip]{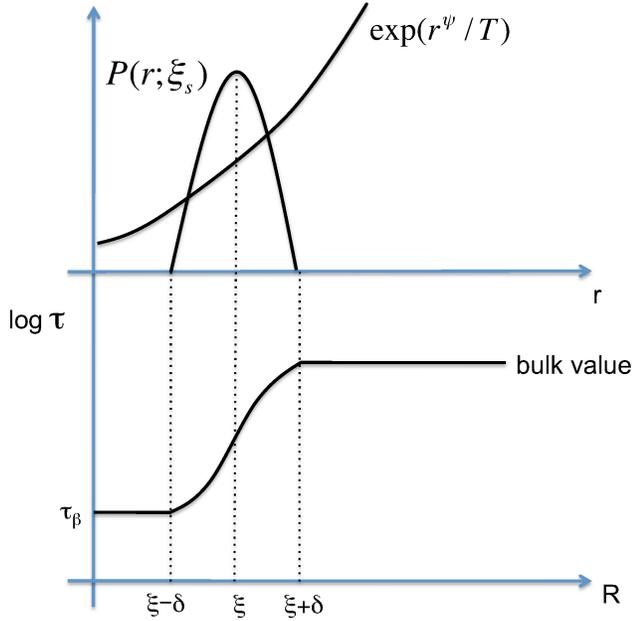}
\caption{Schematic view of the second integral in equation
  \eqref{nasone}. The upper panel represents the two functions within
  the integral, the lower panel is the resulting relaxation time.}
\label{cartoon1}
\end{figure}

To see more precisely how this happens, let us write the surface
tension distribution as $P(Y;Y_c)$, where now $Y$ is the fluctuating
surface tension, whereas $Y_c$ is its typical scale, defined by the
peak of the distribution. This means that a region of radius $R$ will
rearrange or not rearrange, depending on the value of $Y$;
accordingly, its relaxation time can be the either the in-state time
$\tau_\beta$, or the time needed to activately rearrange the region,
\begin{equation}
\tau(R,Y) \sim
  \begin{cases}
    \tau_\beta & Y> T\Sigma R^{d-\theta}  \\
     \tau_0\, \exp\left[\frac{1}{T}\left(Y/T\Sigma\right)^\frac{\psi}{d-\theta}\right]
  & Y< T\Sigma R^{d-\theta}   .
  \end{cases}
\label{sukia}
\end{equation}
The macrosopic equilibration time will be given by an average over $Y$ of the time in \eqref{sukia},
\begin{equation}
\begin{split}
\tau(R) & = \tau_\beta
\int_{T\Sigma R^{d-\theta}}^\infty P(Y;Y_c)\,dY \ + \\
+\ &\quad \tau_0 \int_0^{T\Sigma R^{d-\theta}} P(Y;Y_c)\;
\exp\left[\frac{1}{T}\left(Y/T\Sigma\right)^\frac{\psi}{d-\theta}\right]
\, dY  . 
\end{split}
\label{naso}
\end{equation}
\noindent
The first term in \eqref{naso} corresponds to regions surrounded by
large surface tension, which do not rearrange, and it equals at most
$\tau_\beta$.  The second term corresponds to the low surface tension
regions that do rearrange, and at low temperatures this term is
large. Clearly, if $P(Y;Y_c)=\delta(Y-Y_c)$ we recover the step-like
behavior of $\tau(R)$ described in \eqref{assuppa}.  If, on the other
hand, $P(Y;Y_c)$ is broad, the result is nontrivial.

With a fluctuating surface tension we can still define a typical
mosaic correlation length,
$\xi_s=\left(Y_c/T\Sigma\right)^{1/(d-\theta)}$
\cite{self:nphys08}. This relation suggests an obvious change of
variables useful to recast equation \eqref{naso} in a simpler form:
\begin{equation}
\tau(R)   = \tau_\beta
\int_{R}^\infty P(r;\xi_s)\,dr  + \tau_0 \int_0^{R} P(r;\xi_s)\;
e^{r^\psi/T} \, dr  , 
\label{nasone}
\end{equation}
\noindent
where we are now integrating over all possible sizes of the rearranging regions. 
$P(r;\xi_s)$ is the distribution of sizes, which is of course peaked
on $\xi_s$.  To understand the behavior of the function $\tau(R)$ let
us assume that $P(r;\xi_s)$ has a compact support, being different
from zero only in the interval $r\in [\xi_s-\delta:\xi_s+\delta]$, and
zero elsewhere. We have three regimes of $R$ (see Fig.~\ref{cartoon1}):

i) for $R<\xi_s-\delta$ the first integral in \eqref{nasone} is $1$
and the second integral is $0$, so that $\tau(R) = \tau_\beta$;

ii) for $\xi_s-\delta < R < \xi_s+ \delta$ the weight shifts from the
first to the second integral; because of the exponential, which is
large at low $T$, $\tau$ grows with growing $R$, thus giving rise to a
ramp that brings the relaxation time to a value considerably larger
than $\tau_\beta$;

iii) for $R> \xi_s+\delta$, the first integral is $0$, whereas the
second one has reached its saturation value; to know this value, at
low $T$ we can use the saddle point approximation: the maximum of the
integrand occurs approximately for $r\sim \xi_s$, so that $\tau(R)
\sim \tau_0 \ e^{\xi_s^\psi/T}$. This last quantity is nothing else
than the bulk relaxation time, $\tau_\mathrm{bulk}$.

\begin{figure}
  \includegraphics[angle=-90,width=\columnwidth]{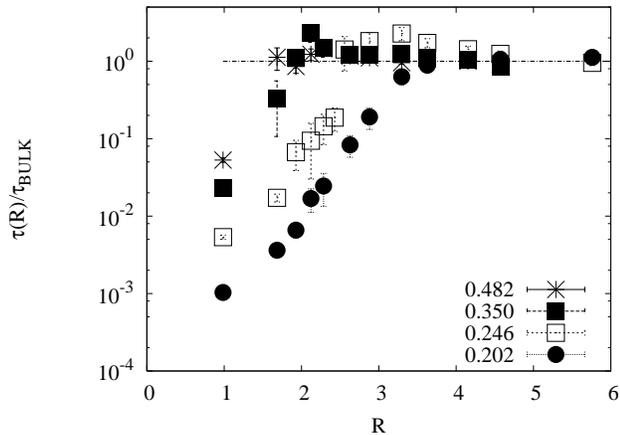}
  \caption{Cavity relaxation time $\tau(R)$, normalized to its bulk
    value, at various $T$. From right to left: $T=0.202, 0.246, 0.350,
    0.482$. The saturation length scale of $\tau(R)$, reported in the
    next figure, is defined by the crossing of the large $R$ (bulk)
    value of $\tau$ and the linear fit (in semi-log) of the small $R$
    data.  This method is straightforward at lower $T$, where
    $\tau(R)$ is unambiguously monotonic, but less so at higher $T$,
    where a small overshooting exists.  We will discuss the origin of
    this overshooting in the following Sections.}
\label{fig:tau-allT}
\end{figure}

What we have just described is a smooth growth of $\tau(R)$ from
$\tau_\beta$ up to the bulk relaxation time $\tau_\mathrm{bulk}$,
taking place in a range of $R$ around $\xi_s$,
\begin{equation}
 \tau(R) \sim 
\begin{cases}
\tau_\beta & \text{for $R < \xi_s-\delta $}  \\
 \text{growth}    & \text{for $\xi_s-\delta < R <\xi_s+\delta$}  \\
 \tau_0\; e^{\xi_s^\psi/T} & \text{for $R > \xi_s+\delta $} \  .
\end{cases}
\label{topa}
\end{equation}
\noindent
It is difficult to better specify the behavior of $\tau(R)$ in its
increasing regime with no knowledge of the distribution $P(r;\xi_s)$
(or, equivalently, of $P(Y;Y_c)$).  Still, in the saddle point limit
(low $T$) there is something we can say: the second integral in
\eqref{nasone} is dominated by the exponential, and for $R<\xi$ the
saddle-point coincides with the right edge of the integration domain,
$r_{\mathrm SP}=R$.  In this case we have,
\begin{equation}
\tau(R) \sim \tau_0\ e^{R^\psi/T},   \qquad \xi_s-\delta < R <\xi_s+\delta  .
\end{equation}
The behavior described by \eqref{topa} is in agreement with what we
have found in our swap simulations (Fig.~\ref{fig:tau149}).  The
relaxation time grows with the radius of the cavity, and it saturates
to its bulk value at $R\sim \xi_s$, so that we can use the saturation
point as an estimate of the static correlation length $\xi_s$.  In
Fig.~\ref{fig:tau-allT} we report the cavity swap relaxation time
normalized by its bulk value for several different temperatures. We
can see that the saturation point moves to larger values of $R$ at
lower temperatures, a phenomenon consistent with the expectation that
the correlation length grows when cooling the system. This fact
consolidates the idea that the point where the cavity relaxation time
saturates is indeed the same static correlation length as extracted
from the point-to-set correlation function.

\begin{figure}
\includegraphics[angle=-90,width=\columnwidth]{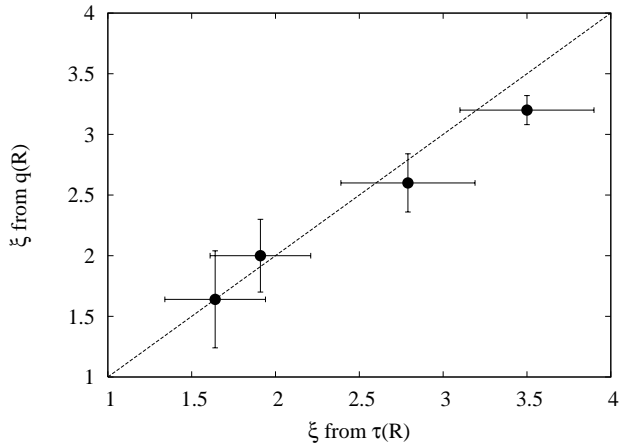}
\caption{Comparison between the correlation length extracted from the
  cavity relaxation time (abscissa) and the point-to-set (PTS)
  correlation length (ordinate).  The PTS correlation length has beed
  defined as the value of $R$ at which the PTS correlation function
  crosses a certain value $\eta$. Given the arbitrariness of $\eta$,
  its value has been chosen in such a way to have the two length
  scales equal at $T=0.482$.}
\label{fig:xi_vs_xi}
\end{figure}

We test this interpretation by plotting in Fig.~\ref{fig:xi_vs_xi} the
length scale of saturation of the swap relaxation time vs.\ the value
of the static correlation length extracted by the point-to-set
correlation function computed in \cite{self:nphys08}. Considering that
both length scales have a degree of arbitrariness in their
measurement, we normalize them in order to be equal at one specific
temperature (see the caption of Fig.~\ref{fig:xi_vs_xi}). Even though
we definitely would need a wider temperature range to say something
certain, we can conclude that the two length scales track each other
quite reasonably. This supports the idea that the point-to-set
correlation length (an eminently static concept) can actually be
measured also by using the swap relaxation time of a cavity subject to
amorphous boundary conditions.

%%%%%%%%%%%%%%%%%%%%%%%%%%%
\section{When cooler is faster}
%%%%%%%%%%%%%%%%%%%%%%%%%%%

\begin{figure}
\includegraphics[angle=0,width=\columnwidth,clip]{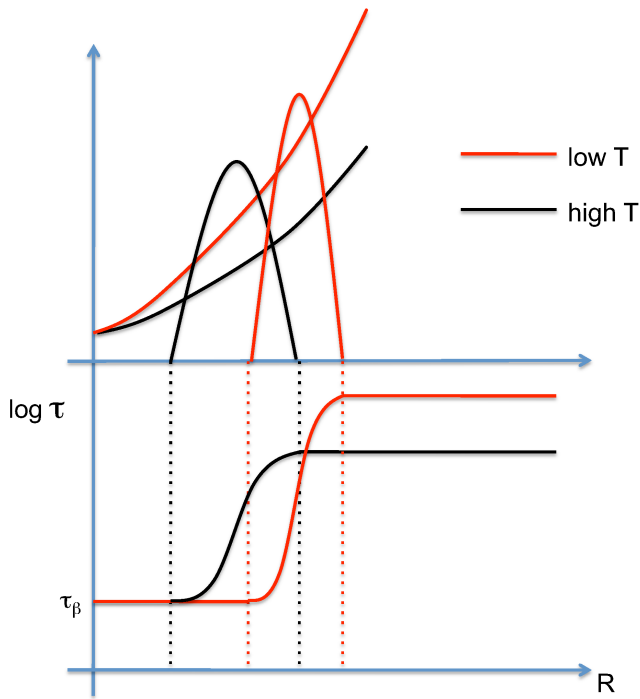}
\caption{Schematic view showing how an inversion of the cold and hot
  relaxation times can take place. By lowering the temperature two
  (related) phenomena occur: i) the correlation length increases, so
  that the distribution $P(r;\xi_s)$ moves overall to the right (it also
  becomes more peaked, see \cite{self:jstatmech09}, but this is
  irrelevant here); ii) the asymptotic bulk relaxation time increases,
  so at saturation $\tau(R)$ reaches a higher level. These two phenomena
  give rise to a regime, between the two correlation lengths, where
  the relaxation time of the colder cavity is lower than that of the
  hotter cavity.}
\label{cartoon2}
\end{figure}

\label{sec:when-cooler-faster}
Both the stepwise behavior of \eqref{assuppa} and the smooth growth of
\eqref{topa} have an interesting consequence: at some values of $R$ a
colder cavity may be {\it faster} than a hotter cavity. How this
happens is pictorially explained in Fig.~\ref{cartoon2}. By lowering
the temperature, $\xi_s$ increases, so we push to the right the support of $P(r;\xi_s)$, and therefore the range
of $R$ over which the growth of $\tau(R)$ occurs; at the same time,
the bulk relaxation time increases, so that the low $T$ curve must
saturate at a higher level than the high $T$ curve. This mechanism
gives rise to a crossing of the cold and hot relaxation times, so that
in the region of $R$ between the cold and hot value of $\xi_s$, we have
that the lower $T$ cavity has a smaller relaxation time than the
higher $T$ cavity.

\begin{figure}
  \includegraphics[angle=-90,width=\columnwidth]{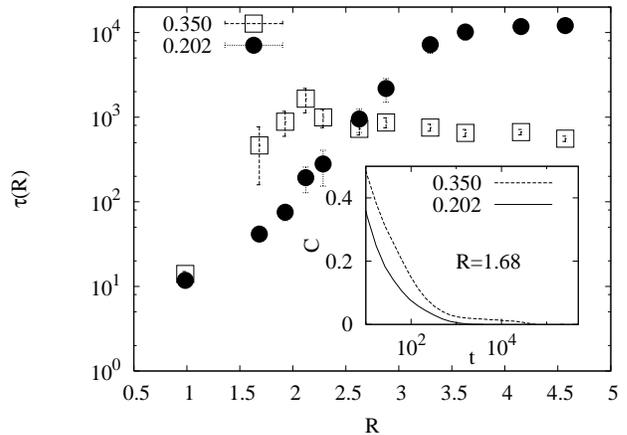}
  \caption{Swap simulations show that an inversion of the relaxation
    time indeed happens: there is an intermediate regime of
    $R\in[1.5,2.5]$ where the low temperature cavity (full circles,
    $T=0.202$) is faster than the high temperature cavity (open
    squares, $T=0.350$). Inset: the autocorrelation function at fixed
    $R=1.68$ at the two different temperatures. Irrespective of the
    definition of the relaxation time, the cooler cavity is faster.}
\label{fig:inversion}
\end{figure}

This odd phenomenon is confirmed by our swap simulations. In
Fig.~\ref{fig:inversion} we show the cavity swap relaxation time at
two different values of $T$. It can be seen quite clearly that for
certain values of the radius the cold cavity is faster than the warm
cavity.  In the inset of Fig.~\ref{fig:inversion} we directly show the
two autocorrelation functions for one specific value of $R$, just to
make clear that the effect does not depend on the particular
definition of $\tau$.

As we have seen, this interesting phenomenon is quite naturally
explained in the context of RFOT. In the sharp scenario, the inversion
of cold and warm relaxation times is a direct consequence of the
presence of two qualitatively different times: the short in-state
time, $\tau_\beta$, and the long out-state relaxation time,
$\tau_\mathrm{bulk}$. The existence of these two times means that at a
certain value of $R$ a cold cavity may still be trapped into its
original state, therefore having a {\it short} in-state relaxation
time, whereas a warm cavity may be unlocked, and therefore have a
longer relaxation time. We remark, once again, that one is comparing
qualitatively different times: the in-state time $\tau_\beta$ is the
time needed to relax within a state, with no cooperative
rearrangement, while the relaxation time of a large cavity,
$\tau_\mathrm{bulk}$ is the time needed for a full rearrangement. Such
distinction is sharp, and easy to detect, only in the stepwise
scenario of equation \eqref{assuppa}. On the other hand, as we have
seen, in the real case $\tau(R)$ (averaged over many samples) is a
smooth function, with a ramp connecting the in-state time to the bulk
time, so that it is harder to distinguish the two different processes
from the full $\tau(R)$ curve. The inversion of cold and hot
relaxation times is therefore an interesting remnant of the presence
of these two different time scales.

%%%%%%%%%%%%%%%%%%%%%%%%%%%%%%%%%%%%%%
\section{An unexpected inequality}
%%%%%%%%%%%%%%%%%%%%%%%%%%%%%%%%%%%%%%

\label{sec:an-unexp-ineq}
In order to have a finite bulk equilibration time, we need the second
integral in equation \eqref{naso} to be finite for
$R\to\infty$. Therefore $P(Y;Y_c)$ must decay sufficiently fast to
suppress the Arrhenius factor. If we make the reasonable assumption,
\begin{equation}
P(Y;Y_c) \sim e^{-(Y/Y_c)^\nu} , \qquad Y \gg 1   ,
\end{equation}
we must have,
\begin{equation}
\nu \geq \frac{\psi}{d-\theta}   .
\label{peppe}
\end{equation}
As we have seen, the distribution $P(Y;Y_c)$ implies an equivalent
distribution of the rearranging regions' size, $P(R;\xs)$,
inequality~(\ref{peppe}) means that $P(R;\xi_s)$ must decay fast
enough to suppress the growth of the equilibration times for large
$R$. This is reasonable.  In \cite{self:nphys08} it was shown that the
exponent $\nu$ is related to the anomaly exponent $\zeta$ that rules
the nonexponential decay of the point-to-set correlation function
$q(R)$,
\begin{equation}
q(R) \sim e^{-(R/\xi_s)^\zeta} ,
\end{equation}
with 
\begin{equation}
 \zeta = \nu(d-\theta) ,
\qquad \zeta \geq 1   .
\end{equation}
where $\theta$ is the surface tension (or stiffness) exponent.  This
leaves us with the inequality,
\begin{equation}
\zeta \geq \psi  .
\label{zanza}
\end{equation}
On increasing the temperature the anomaly $\zeta$ must go to $1$, as
the point-to-set correlation function $q(R)$ becomes a pure
exponential \cite{self:nphys08}. If $\psi$ is temperature-independent,
relation~(\ref{zanza}) then implies,
\begin{equation}
\psi \leq 1   .
\label{luxury}
\end{equation}
We note that the value $\psi\sim 1$ previously reported in
\cite{self:jcp09a} satisfies~(\ref{luxury}). Of course, if we allow
$\psi$ to depend on $T$ (as $\zeta$ does), then there would be no
reason for \eqref{luxury} to be valid in general, whereas
\eqref{zanza} would still hold.

%%%%%%%%%%%%%%%% NONSWAP RESULTS %%%%%%%%
\section{Nonswap dynamics in the confined cavity}
%%%%%%%%%%%%%%%%%%%%%%%%%%%%%%%%%%%%

\label{sec:nonsw-dynam-conf}
\begin{figure}
  \includegraphics[angle=-90,width=\columnwidth]{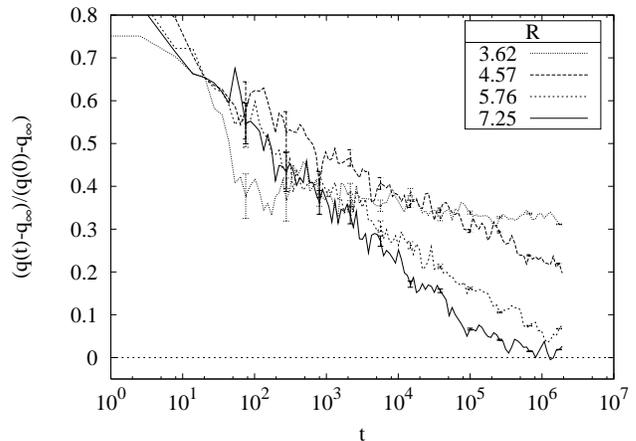}
  \caption{Standard nonswap Monte Carlo dynamics. Connected overlap as
    a function of time for four different sizes of the cavity. The
    connected overlap is obtained by subtracting its equilibrium
    infinite time limit, $q(R)$, obtained with swap, and its
    asymptotic value must be equal to zero. Smaller sizes are
    significantly slower than larger sizes.  For $R\leq 3.5$ the
    dynamics is completely stuck. $T=0.246$.}
\label{fig:nonswap}
\end{figure}

The dynamical behavior of the cavity when we switch off the swap moves
is completely different from what we have seen until now: in contrast
to the swap case, the relaxation is {\it slower} the smaller the
cavity. In the bulk, the dynamics without swap is known to be
significantly slower than with swap \cite{self:pre01} (this is why
swap has been introduced in the first place). However, in the cavity,
not only is nonswap dynamics slower, but the whole dynamical
behavior as a function of $R$ is reversed.

\begin{figure}
  \includegraphics[angle=-90,width=\columnwidth]{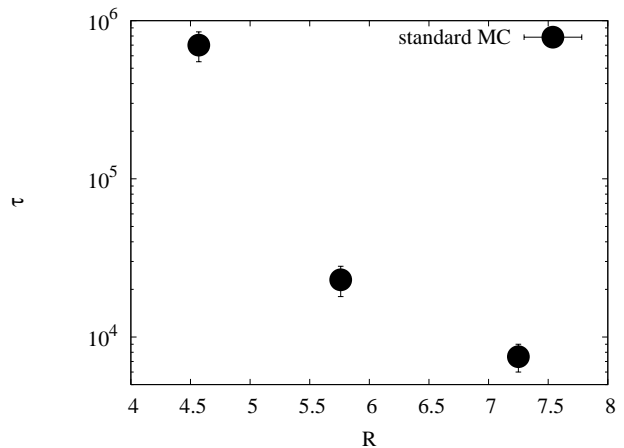}
  \caption{Standard nonswap Montecarlo dynamics. Relaxation time
    obtained by crossing the overlap time series in
    Fig.~\ref{fig:nonswap} with the arbitrary value $\bar q=0.25$.
    For smaller values of $R$ the nonswap dynamics is completely
    stuck, and an extrapolation of $q(t)$ does not make any sense (see
    Fig.~\ref{fig:stuck}).}
\label{fig:taunonswap}
\end{figure}

We observe this phenomenon in Fig.~\ref{fig:nonswap}, where we report
the connected overlap as a function of time in the nonswap case for
different values of $R$. The connected overlap is obtained by
subtracting from $q(t)$ its equilibrium infinite time limit, $q(R)$,
obtained with swap.  The asymptotic value of the connected overlap
must be equal to zero for all $R$ and this makes it easier to compare
different sizes on the same plot. Smaller cavities are dramatically
slower than larger ones. Under these conditions, it is clear that we
cannot compute the overlap autocorrelation function in the nonswap
case, as the system is robustly out of equilibrium. The only time
correlation function that we can use is the overlap itself, $q(t)$,
and to extract a relaxation time, $\tau(R)$, we cross $q(t)$ with an
arbitrary value, $\bar q$.  For those (few) values of $R$ for which
this procedure is viable, we report $\tau(R)$ in
Fig.~\ref{fig:taunonswap}.

In smaller cavities, below $R\sim 4$, the nonswap overlap is
completely stuck to a level which is {\it above} its equilibrium
value.  We can clearly see this by using a BIC (Beta Initial
Condition) test.  The idea is to initialize the cavity in a
configuration $\beta$ which has overlap equal to zero with the
$\alpha$ configuration used to thermalize the system, and which is
frozen in the boundary condition. In this way, the BIC overlap
$q_{\alpha\beta}(t)$ is zero at time zero, and it must increase to the
same asymptotic value as the standard overlap
$q_{\alpha\alpha}(t)$. When thermalization of the cavity is achieved
the two overlaps must meet at the same equilibrium value, $q(R)$
\footnote{This is somewhat similar to the tests introduced by Bhatt
  and Young \cite{spin-glass:bhatt88} and later Katzgraber et al.\
  \cite{spin-glass:katzgraber01} as a thermalization check in
  simulations of spin glasses.}.

\begin{figure}
\includegraphics[angle=0,width=\columnwidth]{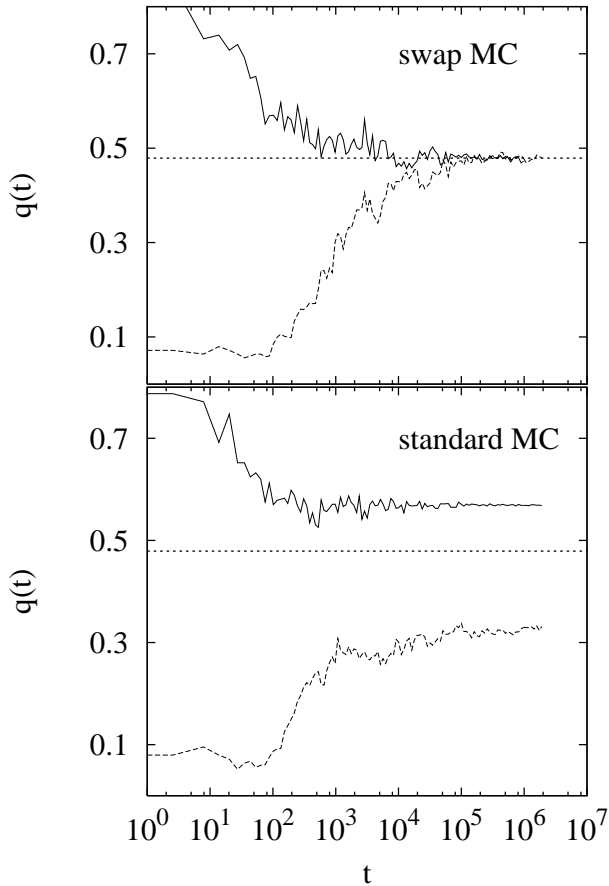}
\caption{BIC test comparison between swap and nonswap dynamics. In the
  BIC test the configuration is initialized both in the same
  configuration $\alpha$ as the frozen boundary (upper, full curve)
  and in a different configuration $\beta$ with respect to the frozen
  boundary (lower, dashed curve). The upper and lower curves must
  reach the same asymptotic value $q(R)$ for infinite times. The BIC
  test is positive for the swap dynamics; all of our swap data, fore
  every value of $R$ and $T$, have passed the BIC test. On the other
  hand, the BIC test is negative for the nonswap
  dynamics. Nevertheless, the nonswap time series is stationary,
  making it impossible to estimate a reasonable value of the
  relaxation time. $R=2.27$, $T=0.246$.}
\label{fig:stuck}
\end{figure}

A positive BIC test is shown in the upper panel of
Fig.~\ref{fig:stuck} for the swap dynamics at small R: the two overlap
branches meet at their asymptotic value, $q(R)$. We have run BIC tests
for all our values of $R$ and $T$ in the swap case, always getting a
positive result (the same holds for the data of
\cite{self:nphys08}). In the lower panel of the same figure we see
what happens in the nonswap case for the same value of $R$: despite
the fact that the overlap is stationary for several decades, it is
definitely not thermalized, as there is a clear and significant gap
between the two branches, none of which reaches the equilibrium value
$q(R)$ (dotted line). Hence, at this value of $R$ and of $T$ it is not
even possible to roughly estimate $\tau$: no extrapolation of $q(T)$,
however wild, makes sense with these data.

We notice that this slowing down happens also at relatively high
temperatures: the effect of the confinement on the relaxation time is
really drastic, and the difference between swap and nonswap dynamics
stark. Incidentally, we note that without swap dynamics it would be
impossible to measure the point-to-set correlation function (which is
the equilibrium value of $q$), due to this hyper-slowing down. The
slowing down of the dynamics in a confined cavity was noted before in
ref.~\cite{confinement:berthier11} for molecular dynamics.

In the next Sections we address the conflict between the swap and
nonswap results.

%%%%%%%%%%%%%%%%%%%%%%%%%%%%%%%%%%%%%%
\section{The contradiction between swap and nonswap}
%%%%%%%%%%%%%%%%%%%%%%%%%%%%%%%%%%%%%%

\label{sec:contr-betw-swap}

At this point we are left with a contradictory scenario. On one hand,
with swap Monte Carlo the relaxation time grows up to its bulk value
when increasing the cavity radius $R$, seemingly saturating when $R$
reaches the point-to-set correlation length $\xi_s$. This behavior
suggests that $\xi_s$ is indeed the typical size of the cooperatively
rearranging regions, which dominate activated dynamics at low
temperatures. On the other hand, with standard nonswap Monte Carlo (as
well as molecular dynamics \cite{confinement:berthier11}), the cavity
relaxation time is larger than its bulk value and grows with
decreasing $R$.

A dramatic increase of the nonswap relaxation time might suggest some
kind of phase transition. Indeed a scenario involving a true phase
transition has been recently described in
ref.~\cite{mosaic:cammarota11}.  However, an essential ingredient of
any phase transition is the thermodynamic limit. There is no true
divergence at finite volume, but rather an unbounded growth of the
relaxation time with volume. The transition discussed in
ref.~\cite{mosaic:cammarota11} applies to geometries where it is
possible to send the system size to infinity (for example, scattered
frozen particles or a sandwich geometry---see
ref.~\cite{confinement:berthier11}), in which case the relaxation time
for $R\sim\xi_s$ should diverge. However, in our cavity geometry, the
size is always finite, so that a phase transition does not seem the
right explanation of what we see.  So the nonswap scenario is hard to
interpret within an RFOT context and it seems to be more compatible
with a theory of the glass transition based on the idea that dynamics
is facilitated by defect diffusion \cite{review:chandler10}: the
smaller the volume, the smaller the number of defects and the slower
the dynamics.

But even if the nonswap dynamics results would seem more physically
relevant, a complete physical picture needs to account also for the
swap results (in particular the intriguing fact of the saturation of
$\tau$ at $R\sim\xi_s$) and to resolve the contradiction. This is what
we attempt, in a rather speculative way, in this Section.

\subsection{The hybridization between MCT and activation}

\subsubsection{The bulk case}

To better understand what is going on in the cavity, we have to go
back to the bulk. According to some theories of the glass transition
\cite{review:biroli09}, there are two relaxation channels: a
nonactivated channel, well described by Mode Coupling Theory (MCT)
\cite{rev:goetze92}, which is ruled by unstable stationary points of
the potential energy (saddles), and a second channel, consisting of
activated barrier crossing. The first mechanism has a singularity at
the MCT transition temperature $T_c$, where the MCT relaxation time
diverges as a power law. On the other hand, the activated channel is
insensitive to $T_c$, and its relaxation time increases in a
super-Arrhenius fashion, due the the low-$T$ increase of the static
correlation length, $\xi_s$.

\begin{figure}
  \includegraphics[angle=0,width=\columnwidth,clip]{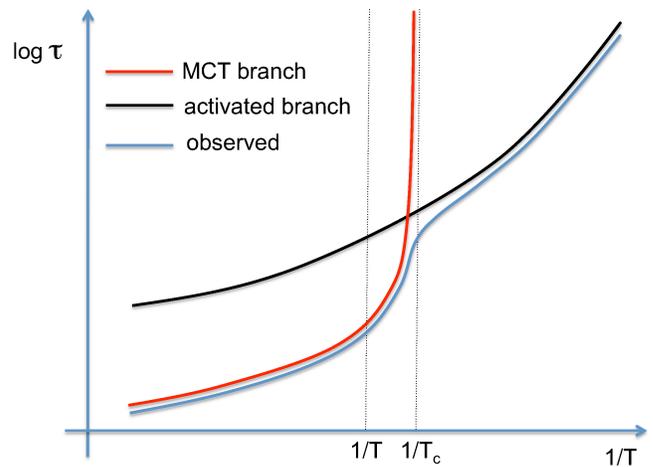}
  \caption{A schematic view of bulk relaxation. We hypothesise that
    there are two different channels of relaxation: i) the Mode
    Coupling Theory (MCT) channel, which is related to a relaxation
    which uses unstable stationary points (saddles) of the potential
    energy.  The MCT dynamics has a relaxation time that diverges at
    $T_c$. ii) the activated barrier-crossing channel. The actual
    dynamics ``chooses'' the fastest of the two channels, so that the
    observed relaxation time is the lowest of the two. Below $T_c$,
    there is a dynamical crossover between the MCT branch to the
    activated branch. The crossover is exaggerated here to illustrate
    the point, the actual behavior is much smoother in the $T\sim T_c$
    region.}
\label{cartoon3}
\end{figure}

We make the hypothesis that the real (observed) relaxation time of the
system is the lowest of the two relaxation times, because the dynamics
always follows the fastest relaxation channel. We can then get an
impression of what happens in Fig.~\ref{cartoon3}. The observed time
follows the MCT branch up to close to $T_c$, where it crosses over to
the activated branch, thus avoiding the MCT divergence. This
hybridization between MCT and activated branches is (very roughly
speaking) the origin of the dynamical crossover near $T_c$
\cite{review:biroli09}.

\begin{figure}
  \includegraphics[angle=0,width=\columnwidth,clip]{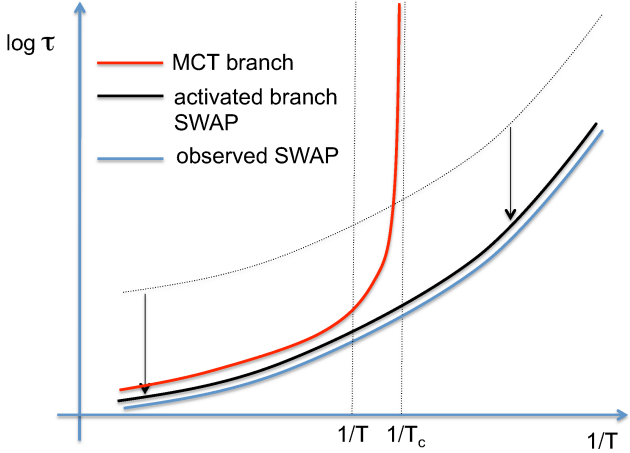}
  \caption{When we use a swap dynamics we are significantly lowering
    the prefactor of activated barrier crossing, hence shifting
    the activated branch downwards. As a result, there is no
    significant hybridization between the two branches and the
    resulting (observed) swap relaxation time does not detect any
    particular crossover close to $T_c$.}
\label{cartoon4}
\end{figure}

Consider now what happens to this scenario when we use a swap
dynamics. In general the activated relaxation time can be written as,
\begin{equation}
\tau_\mathrm{ACT} = \tau_0 \exp(\xi_s^\psi/T)  ,
\label{belladonna}
\end{equation}
where $\xi_s$ is the static correlation length. The effect of swap
dynamics is essentially to decrease significantly the prefactor
$\tau_0$ in equation \eqref{belladonna} \cite{glassthermo:fernandez06},
\begin{equation}
\tau_\mathrm{ACT}^\mathrm{swap} = \tau_0^\mathrm{swap} \exp(\xi_s^\psi/T)   ,
\qquad \mbox{with }\tau_0^\mathrm{swap} \ll \tau_0   .
\label{bellaswap}
\end{equation}
This amounts to a downward shift of the activated branch
(Fig.~\ref{cartoon4}). Due to this, the hybridization between the two
branches disappears, and the observed relaxation time does not display
any significant crossover close to $T_c$. We also see that if we fix a
temperature $T \gtrsim T_c$, in the nonswap case the bulk time is
dominated by the MCT channel, whereas in the swap case it is dominated
by the activated channel (also see Fig.~1 of
ref.~\cite{glassthermo:fernandez06}, which shows how the MCT plateau
seen in time correlation functions is lost with swap dynamics).

\subsubsection{The cavity}

Let us now turn to the cavity, bearing in mind that the large $R$
value of $\tau(R)$ is nothing else than the bulk time, whose behavior
we have just examined.  It has been suggested that the MCT cavity
relaxation time, as a function of $R$, should have a divergence at
$R\sim \xi_d$, where $\xi_d$ is the {\it dynamic} correlation length
\cite{mosaic:franz07}. A possible interpretation of this fact is that
in a smaller cavity the frozen boundary conditions stabilize unstable
saddles, thus increasing the MCT relaxation time. Below $\xi_d$ the
cavity runs out of saddles and nonactivated relaxation becomes
impossible. On the other hand, the activated relaxation time obeys the
scenario described by eqs.~\eqref{nasone} and \eqref{topa}: it
increases with $R$, saturating at the {\it static} correlation length,
$\xi_s$.

\begin{figure}
\includegraphics[angle=0,width=\columnwidth,clip]{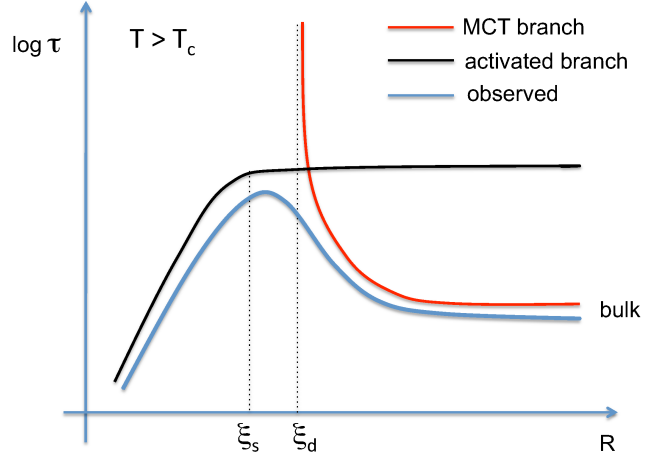}
\caption{In the cavity, for $T\gtrsim T_c$ the hybridization between
  MCT and activated branch may give rise to a nonmonotonic
  $\tau(R)$. Starting from large cavities, the relaxation time follows
  the MCT branch, which has a divergence at the dynamical correlation
  length, $\xi_d$. In the proximity of such divergence the observed
  $\tau(R)$ crosses over to the activated branch, thus decreasing with
  decreasing $R$.}
\label{cartoon5}
\end{figure}

As for the bulk, we can speculate that the observed relaxation time in
the cavity will be the smallest of the two times. Let us fix a
temperature slightly above $T_c$, so that the nonswap bulk relaxation
is dominated by the MCT channel (Fig.~\ref{cartoon3}). In
Fig.~\ref{cartoon5} we get a picture of what happens. Let us start
from large values of $R$: the relaxation time follows the MCT branch,
therefore giving an {\it increase} of $\tau(R)$ for decreasing
$R$. But at some point the MCT branch crosses the activated one (and
it eventually diverges at $\xi_d$), so beyond this point the dynamics
sticks to the activated channel, giving rise to a maximum of
$\tau(R)$. Hence, for small values of $R$ we recover a regime where
$\tau(R)$ decreases for decreasing $R$.

The large $R$ regime of this nonmonotonic curve was also discussed in
\cite{review:biroli09}, where it was noted that above $T_c$ $\tau(R)$
should approach its bulk value from above. This behavior, namely a
relaxation time that increases from its bulk value when decreasing
$R$, is indeed what we find with nonswap dynamics,
Fig.~\ref{fig:taunonswap}.  However, in the nonswap case the increase
of the relaxation time is so sharp that we struggle to follow this
curve down to medium-small $R$, so we cannot access the overshooting.

\begin{figure}
\includegraphics[angle=0,width=\columnwidth,clip]{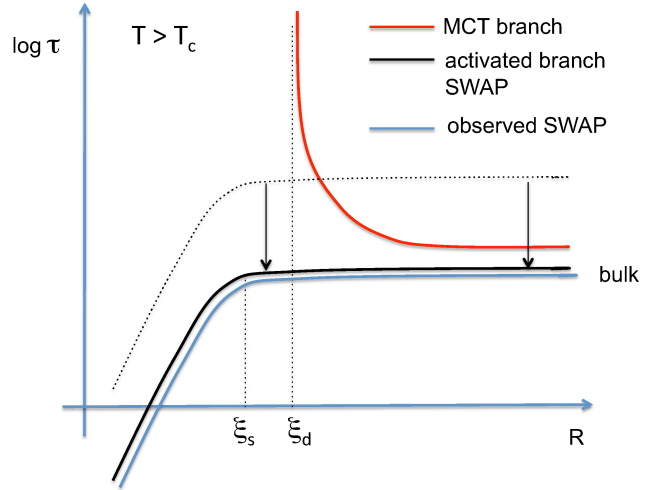}
\caption{When we use swap dynamics in the cavity we shift
  the whole activated branch downwards, hence lowering the degree of
  hybridization of the two branches. In this way, the overshooting of
  relaxation time may be completely washed out, and $\tau(R)$ have a
  purely monotonously increasing behavior.}
\label{cartoon6}
\end{figure}

\begin{figure}
  \includegraphics[angle=-90,width=\columnwidth]{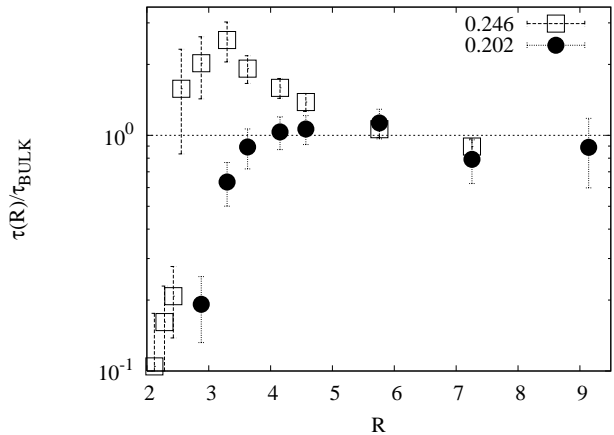}
  \caption{Cavity relaxation time with swap dynamics. This is a
    zoom-in of Fig.~\ref{fig:tau-allT}, made to emphasize the
    nonmonotonic behavior of $\tau(R)$.  At the highest temperature
    (open squares, $T=0.350$) there is an overshooting of $\tau(R)$,
    caused by the hybridization between the MCT and activated
    branches. At the lowest temperature (filled circles, $T=0.202$)
    the overshooting disappears due to the decreased gap between the
    MCT and the activated branch. The relaxation times are normalized
    by their bulk value.}
\label{fig:overshooting}
\end{figure}

What happens when we use swap dynamics? As in the bulk, by using swap
we are decreasing the prefactor of activation, thus shifting the whole
activated branch downwards. From Fig.~\ref{cartoon6} we see that this
shift has the effect to weaken, or even wash out entirely, the
nonmonotonic behavior of $\tau(R)$. Something similar happens by
lowering the temperature (getting closer to $T_c$), because in that
way we are narrowing the difference between the MCT and the activated
branch (Fig.~\ref{cartoon3}). In the cavity, this amounts to closing
the gap between the two branches at large $R$.  Hence, we expect that
lowering $T$ too has the effect to iron out the maximum of $\tau(R)$,
eventually making it disappear \footnote{ This is a general prediction
  of our picture: by lowering the temperature we are gradually pushing
  up (and therefore ruling out) the MCT branch, diminishing the
  hybridization of the two branches and therefore eliminating the
  overshooting. At very low $T$, $\tau$ should be a purely increasing
  function of $R$.}.  Summarizing, we expect swap dynamics to display
little sign of a nonmonotonic cavity relaxation time $\tau(R)$, and to
become completely monotonic at low $T$.  In
Fig.~\ref{fig:overshooting} we show a close-up of the cavity
relaxation time with swap dynamics at two different temperatures:
there is an overshooting of $\tau(R)$ at intermediate temperature, but
it completely disappears a the lowest $T$. This expectation is
therefore supported by the data.

According to this scenario (admittedly based on little evidence), in
the nonswap case one should see an increase of $\tau(R)$ over its bulk
value when decreasing the cavity size from large $R$ (saturation from
above), whereas in the swap case (and at low $T$) the cavity
relaxation time should decrease below its bulk value when decreasing
the radius (saturation from below). This prediction seems to be in
qualitative agreement with our numerical findings. However, there is a
severe problem with this interpretation, namely the fact that nonswap
dynamics at very small $R$ is stuck.  Even assuming that the great
increase of the cavity relaxation time that we observe in going from
large $R$ down to medium $R$ (Fig.~\ref{fig:nonswap}) has to be
identified with the large $R$ regime of a nonmonotonic $\tau(R)$
(Fig.~\ref{cartoon5}), the question remains: why we do not see any
hint of the low $R$ regime of Fig.~\ref{cartoon5}, where the cavity
relaxation time gets smaller for small radii?  It is well possible, in
this scenario, that for intermediate $R$ the relaxation time is
significantly larger than the bulk limit. However, for {\it very}
small $R$ the relaxation time should drop again. Yet, we do not see
this. In fact, very small cavities are {\it completely stuck}, as
shown in Fig.~\ref{fig:stuck}. This phenomenon is in open disagreement
with our theoretical expectation. We must address this inconsistency.

\subsection{The role of boundary rearrangements}

The fact that swap dynamics thermalizes a small cavity quite rapidly
while nonswap dynamics remains stuck, is weird; it indicates that
swapping different particles in a small volume becomes prohibitive for
standard dynamics. Of course, the exchange of two particles for the
standard dynamics is the result of many moves, and it is for sure a
more complicated process. Yet swapping two nearby particles is
definitely not a terribly collective rearrangement and it should not
implicate a very large activation barrier. If it does, it means that
this barrier has been made dramatically large by the amorphous
boundary conditions. Why is that?

A possible explanation is that by freezing the external configuration
we are preventing the surrounding system to elastically accommodate
for the small rearrangements within the cavity. Although exchanging
two particles is not a collective rearrangement, i.e.\ one in which
many particles move a lot, to happen it still needs that many
particles make very small movements. This phenomenon was studied in
ref.~\cite{self:prl02}, where the distribution of particle
displacements in moving from a local energy minimum to nearby one
connected by a saddle of order $1$ was calculated. It was found that
this process corresponds to {\it few} particles (order 2--3) moving an
amount comparable to the interparticle distance and {\it many}
particles moving very little, just to make space to the rearranging
ones. Elasticity is also a central ingredient in the local elastic
expansion model (also called ``shoving model'') of viscous
relaxation~\cite{dynamics:dyre96}. More in general, one might argue
that the whole short-time dynamics (not only elastic modes) plays a
relevant role.

By freezing all the particles in the configuration external to the
cavity we are inhibiting this contribution, perhaps making unnaturally
large an otherwise modest barrier. Swap dynamics, on the other hand,
needs not to pass through the top of a barrier to exchange two
particles, and therefore is less affected by the suppression of the
high-frequency response, and by the subsequent barrier's
increase. This may be the origin of the very different qualitative
behavior of swap vs. nonswap dynamics observed at low $R$.

To check this last hypothesis, we suggest in the next Section a
general approach to restore the short-time dynamics which is not
limited to the elastic case, suggested by an alternative description
of the problem in the RFOT spirit.

%%%%%%%%%%%%%%%%%%%%%%%%%%%%%%%%%%%%
\section{Frozen configuration vs. frozen state}
%%%%%%%%%%%%%%%%%%%%%%%%%%%%%%%%%%%%

\label{sec:fc-vs-vs}

An alternative description of the over-constraining due to the
boundary can be given in terms of configurations vs.\ states. The
original aim of the amorphous boundary conditions
\cite{mosaic:bouchaud04} was to keep the system surrounding the cavity
within one fixed {\it state} (say $\alpha$), one of the exponentially
many metastable states the supercooled liquid phase is composed of
\cite{mosaic:kirkpatrick89}. According to this spirit, the external
particles should be allowed to move enough to visit the many
configurations belonging to state $\alpha$, but not enough to reach
configurations that do not belong to $\alpha$. By choosing and fixing
just one {\it configuration} within state $\alpha$, however, we are
over-constraining the amorphous boundary, and this may have some
side-effects on the dynamics of small rearrangements in the cavity
when a standard dynamics is used.

In view of this, it seems reasonable to try to relax the constraint on
the outer particles by changing the current {\it frozen configuration}
(FC) setup, in favour of a {\it frozen state} (FS) one. This means
that instead of completely freezing the particles outside the sphere,
we let them relax subject to the condition that the overlap
$q_\mathrm{ext}$ between the initial external configuration and the
one at time $t$ remains at some value $\hat q_\mathrm{ext}$. The FC
setup would be recovered taking simultaneously the limits
$q_\mathrm{ext}\to1$ and $\ell\to0$. In this way, the external
configuration is not allowed to move at all, so this amounts to a
complete freezing\footnote{We remark though that the FC results
  reported above are in fact obtained with \emph{bona fide}
  freezing.}.

Of course, the choice of $\hat q_\mathrm{ext}$ is critical: with too large
a value we go back to the frozen configuration case, while too small a
value destroys any point-to-set correlation in the cavity. In fact, in
the limit $\hat q_{\mathrm ext}= 0$ the cavity must be ergodic and the
overlap must relax to zero for any value of $R$. A sensible physical
choice is,
\begin{equation}
\hat q_{\mathrm{ext}} = q_{\mathrm{EA}}  ,
\label{FSbotto}
\end{equation}
where $q_{\mathrm EA}$ is the self-overlap of a metastable state. In
this way we ensure that the external system does not make any major
structural rearrangement, and yet allows for minor movements of the
particles, which can have an important elastic effect. To chose the
correct value of $q_\mathrm{EA}$ we use the thermodynamic potential
$V(q)$ recently discussed in \cite{self:prl10}, whose secondary
minimum indicates the value of the self-overlap $q_{\mathrm EA}$. At
the temperature $T=0.246$, where we will run the FS simulations, a
reasonable choice is (see Fig.~5 of ref.~\cite{self:prl10}),
\begin{equation}
q_{\mathrm EA} = 0.4  .
\end{equation}
Of course, the final test for this choice is that the point-to-set
correlation must not be lost: having switched from a FC to a FS setup
will certainly imply that the infinite time limit of the overlap,
$q(R)$, will be smaller at all values of $R$. What we need is
$q(R)\neq 0$ at for some range $[0:R]$, in order to have a physically
significant point-to-set correlation function.

%%%%%%%%%%%%%%%%%%%%%%%%%%%%%%%%%%%
\section{Cavity dynamics with frozen state boundary conditions}
%%%%%%%%%%%%%%%%%%%%%%%%%%%%%%%%%%%

\label{sec:cavity-dynamics-with}

\subsection{FS simulations: technical details}

First, we need to make a technical, but relevant, remark. In this
work, as well as in previous works \cite{self:prl07,self:nphys08}, the
overlap is defined in such a way that it \emph{does not} detect the
exchange of particles of different size. The same definition has been
adopted by other groups \cite{confinement:berthier11,
  mode-coupling:berthier11}. However, we cannot use this definition
for imposing the constraint on the external particles: an exchange of
two different particles, perhaps quite far from each other, must not
be allowed. Hence, the constraint must be imposed on an overlap that
is sensitive to the exchange of particles of different kind (whereas
we still do not distinguish the exchange of identical particles). Let
us call this the {\it binary} overlap, defined as
\begin{equation}
  \label{eq:binary_overlap_def}
  q_\mathrm{bin}(t) \equiv \frac{1}{\ell^3 \, N_i} \sum_{i \in
    v}  \left[ n_i^A(0)n_i^A(t) + n_i^B(0)n_i^B(t)\right] ,
\end{equation}
where $n_i^X(t)$ is the number of particles of kind $X$ in box
$i$. This is also the definition used in \cite{self:prl10} to compute
the thermodynamic potential $V(q)$. In what follows we thus use
$q_\mathrm{bin}$ to put the constraint on the outside particles. On
the other hand, in order to compare with the previous results, we
continue using the standard overlap within the cavity.

Conceptually, FS simulations are straightforward: we simply reject all
moves on the external particles that violate constraint
\eqref{FSbotto}.  In practice, FS simulations are much more demanding
than FC ones, because now we have to update {\it all} particles in the
system, not simply those within the cavity\footnote{Note that also
  in the FS setup, as in the FC one, we use a hard wall potential
  enclosing the particles within the cavity. In this way, particles
  cannot cross the surface of the cavity: whoever is in, stays in, and
  whoever is out, stays out.  This procedure is essential in order to
  obtain the correct thermodynamic ensemble.  }. For this reason we
restricted our investigation of the frozen state setup to just 3
cavity sizes, $M=20, 50, 100$ particles, corresponding to $R= 1.68,
2.27, 2.88$, and to just one temperature, $T=0.246$.

\subsection{FS simulation results}

The first thing we have to check is what happens to the point-to-set
correlation function, i.e.\ to the asymptotic value of the overlap,
$q(R)$, in the FS setup at this value of the temperature. To do this
we run a swap BIC test in the FS setup, to be sure to get the
thermalized asymptotic overlap.  We report these values in Table I,
where we also report the corresponding values for the standard FC
setup. Recall that the effective zero of the overlap, i.e.\ the value
it has for two uncorrelated configurations, is $q_0 = 0.062876$.

\begin{table}[ht]
\caption{Point-to-set correlation function $q(R)$: FS vs. FC}   % title of Table
\centering % used for centering table
\begin{tabular}{c c c c} \\% centered columns (3 columns)
\hline\hline %inserts double horizontal lines
$M$ & $R$ & $q^\mathrm{FS}(R)$ & $q^\mathrm{FC}(R)$  \\ [1ex] % inserts table
%heading
\hline % inserts single horizontal line
20 \quad & \quad 1.68 \quad & \quad 0.222     $\pm$  0.004 \quad &\quad  0.578     $\pm$  0.001  \\ % inserting body of the table
50 \quad& \quad  2.27 \quad & \quad 0.142     $\pm$ 0.003 \quad&   \quad0.479    $\pm$ 0.001  \\
100 \quad&\quad 2.88 \quad & \quad 0.095   $\pm$  0.002 \quad&  \quad 0.314  $\pm$  0.002  \\ [1ex] % [1ex] adds vertical space
\hline %inserts single line
\end{tabular}
\label{zoccola} % is used to refer this table in the text
\end{table}

As expected, there is a significant decrease of $q(R)$ in the FS case,
due to the fact that particles in the external configuration are now
partly free to move, hence lowering the constraint on the inner
particles. However, in the FS case $q(R)$ is still nonzero, so that
the PTS correlation function is nontrivial. We stress that the values
in Table I have been obtained from a swap BIC test: the lower branch
of the BIC test {\it grows} with time up to its asymptotic limit. We
are therefore quite sure that the FS values of $q(R)$ that we report
are nonzero.

\begin{figure}
  \includegraphics[angle=0,width=\columnwidth]{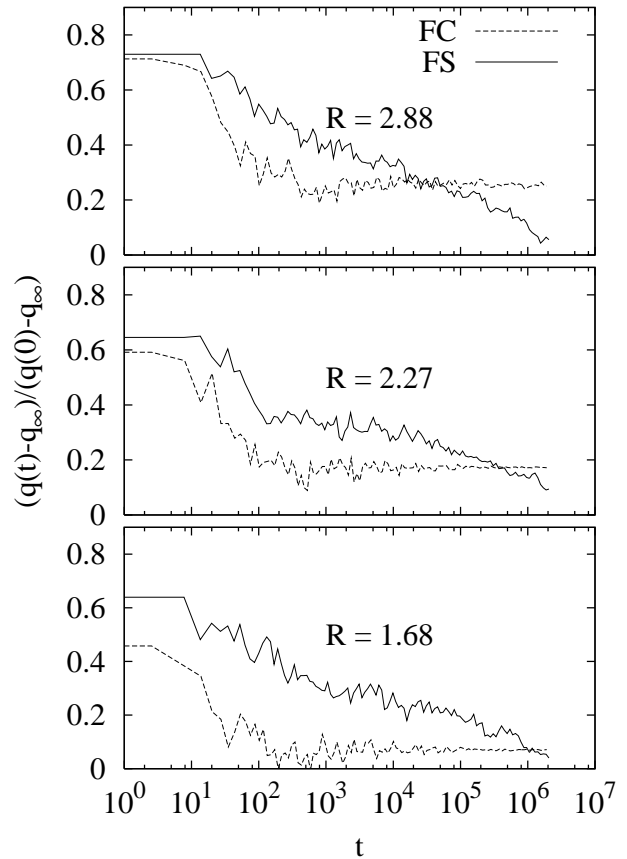}
  \caption{Frozen configuration (FC) vs.\ frozen state (FS) setup,
    standard nonswap dynamics. We plot the connected overlap, obtained
    by subtracting its equilibrium infinite time limit $q(R)$
    (obtained with a swap BIC test). The asymptotic equilibrium value
    of the connected overlap is zero. The three values of $R$
    investigated here are small, so that the FC dynamics (dashed line)
    is completely stuck at an out of equilibrium level. On the
    contrary, the FS dynamics (full line) is not stuck and, even
    though longer runs would be needed, it is approaching
    equilibrium. $T=0.246$.}
\label{fig:punch}
\end{figure}

Next, we turn to the time series of the overlap $q(t)$ in the FS
setup, compared to the FC setup. We stress, once again, that we are
using standard nonswap Montecarlo dynamics, both for FC and FS. The
data are reported in Fig.~\ref{fig:punch} for three different values
of $R$. In order to make the FS/FC comparison easier, we plot the
connected overlap, i.e. the overlap subtracted by the (swap)
equilibrium value, $q(R)$. The connected overlap must go to zero for
infinite time.

At these values of $T$ and $R$, the FC time series (dashed lines) are
completely stuck at an off-equilibrium value, so much as to make it
impossible to even estimated the relaxation time. We already observed
this phenomenon in Fig.~\ref{fig:stuck}.  On the other hand, the FS
time series (full lines) are quite different: the overlap does not
remain stuck at any specific level; in fact, it seems to be decaying
steadily towards zero. Unfortunately $2$ million Monte Carlo steps
(our largest time) are not enough to directly observe the point where
the connected overlap goes to zero. However, a reasonable
extrapolation suggests that for all three values of $R$ the relaxation
time is somewhere between $10^6$ and $10^7$ Monte Carlo steps.

\begin{figure}
  \includegraphics[angle=0,width=\columnwidth]{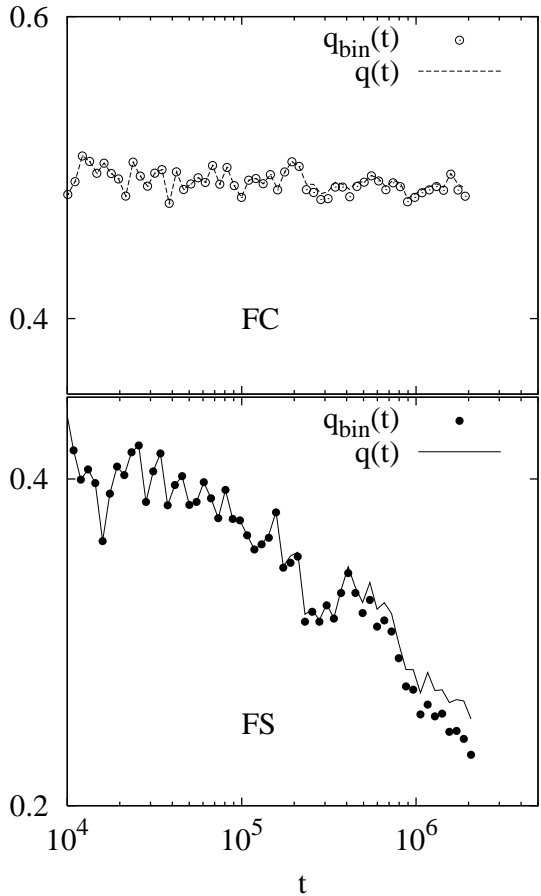}
  \caption{Frozen configuration (FC) vs. frozen state (FS) setup,
    standard nonswap dynamics. We plot the standard overlap (lines)
    and the binary overlap (symbols). These two overlaps are the same
    as long has no particles of different size have been swapped. On
    the other hand, when the (nonswap) dynamics starts swapping
    particles, the binary overlap gets smaller than the standard
    one. This never happens in the FC setup (upper panel), whereas it
    happens for sufficiently long times in the FS setup (lower
    panel). This fact explains why the FC dynamics is stuck, while the
    FS one is not. $R=2.88$, $T=0.246$.}
\label{fig:qqbin}
\end{figure}

We conclude that the cavity dynamics with frozen state boundary
condition no longer remains stuck at an off-equilibrium level. This
result goes in the direction we expected: allowing the in-state
vibrations of the external configuration unleashes some minor, but
necessary, relaxation modes that are otherwise frozen in the FC setup.
In particular, what happens is that in the FS case even a nonswap
dynamics is able (after a while) to exchange different particles,
while in the FC case this never happens. This phenomenon is shown in
Fig.~\ref{fig:qqbin}. We report in this figure the standard cavity
overlap, $q(t)$, and the binary cavity overlap, $q_\mathrm{bin}(t)$,
which is sensitive to the exchange of different particles. What we see
is that in the FC setup (upper panel) the two overlaps coincide up to
the longest time, meaning that particles exchanges never
happen\footnote{Strictly, this means that exchanges of particles
  \emph{of different kind} do not happen, but same-kind exchanges
  should be similarly hindered}. On the other hand, in the FS setup
(lower panel) there is a decoupling between the two overlaps at about
$5\times 10^5$ Monte Carlo steps (this decoupling is also found in FC
swap dynamics ---not shown--- where it is naturally expected since the
swap moves consist precisely in the exchange of two particles of
different kind). Hence, even the nonswap dynamics is able to swap
particles in the long run, and therefore to relax the cavity, provided
that we confine the external system within a state, rather than a
configuration.

The last, and most important, open issue is the behavior of the
relaxation time as a function of $R$. We recall here the situation
schematically summarized in Fig.~\ref{cartoon5}: the relaxation time
for medium $R$ can be significantly larger that the bulk time, due to
the hybridization of the MCT and activated branches. However, for
small enough $R$ one should go back to a regime where $\tau$ decreases
for decreasing $R$. This overshooting scenario is what happens with a
local swap dynamics, Fig.~\ref{fig:overshooting}, and our expectation
was that it should also happen with a normal nonswap dynamics,
provided that we use an FS setup. Is this scenario confirmed or
disproved by the data in Fig.~\ref{fig:punch}?

Longer simulations (at least one order of magnitude longer) and
several more values of $R$ and $T$ would be required to clear up
unambiguously this matter. However, an unscrupulous extrapolation of
the data in Fig.~\ref{fig:punch} suggests that the intermediate sized
cavity, $R=2.27$, has the largest relaxation time, definitely closer
to the right side of the $[10^6:10^7]$ window, whereas the smallest
and largest cavities, $R=1.68$ and $R=2.88$, both seem to have a
smaller relaxation time, closer to the $10^6$ side. In other words,
the smallest cavity seems {\it not} to be the slowest one. If this
were true, it would mean that we are exactly around the maximum of
$\tau(R)$ in Fig.~\ref{cartoon5} and that we are starting to see a
hint of the expected decrease of $\tau$ with decreasing $R$.  Needless
to say, we cannot push this interpretation of the data too far. Let us
be content to say that current simulations with frozen state boundary
conditions and nonswap dynamics do not rule out the existence of a low
$R$ regime where smaller cavities have smaller relaxation times.

Further work to clear up this issue is currently in progress.
Unfortunately there is no easy way to settle this. We cannot use
smaller cavities, because $M=20$ particles is already about the
smallest reasonable size in three dimensions. The only thing to do is
to push the simulations at longer times, which is computationally very
costly.

%%%%%%%%%%%%%%%%%%%%%%%%%%%%%%%%%%%%%
\section{Some experimental evidence}
%%%%%%%%%%%%%%%%%%%%%%%%%%%%%%%%%%%%%

\label{sec:some-exper-evid}

At the experimental level, there has been considerable interest in
studying liquids in confinement conditions, in particular since
nanoporous materials with well-defined pore radius have been available
(see \cite{confinement-exp:liu89, confinement-exp:zhang92,
  confinement-exp:arndt97} and references therein) and more recently
materials such as carbon nanotubes \cite{review:rasaiah08}.

For liquids confined in nanopores the experimental glass transition
temperature (as measured with differential scanning calorimetry) is
\emph{reduced} as the pore becomes smaller
\cite{confinement-exp:zhang92}, i.e.\ confined systems are
\emph{faster} than the bulk. However, the situation is rather more
complicated, as relaxation experiments \cite{confinement-exp:liu89,
  confinement-exp:arndt97} point to the existence of at least two
regions in space, with different dynamics: a \emph{slow} layer of
molecules directly in contact with the pore walls and a \emph{fast}
region inside the cavity and far from the walls.

A particularly interesting case is reported in
ref.~\cite{confinement-exp:arndt97}: the relaxation time of salol
confined in nanoporous silica glass was found with
dielectric relaxation measurements (unable to distinguish the
interfacial and central regions of the pore) to be larger for
increasing confinement. However, after coating the pore walls with a
hydrophobic lubricant (thus reducing the H bonds between salol and the
pore surface) it was found that {\it smaller} cavities are {\it
  faster}. In particular, they are significantly faster than the
bulk. Hence, in this experimental case, once the interactions that
slow down the interfacial layer were supressed, the relaxation time as
a function of the radius has a qualitative behavior similar to
Fig.~\ref{fig:tau149}. The authors of
ref.~\cite{confinement-exp:arndt97} used this to determine a
cooperativity length scale.

There are intriguing similarities, as well as obvious differences,
with our case. In both cases the original interaction with the cavity
interface was too stiff, suppressing some relaxation channels that are
not cooperative, and yet necessary to equilibrate the cavity. The
strategy in \cite{confinement-exp:arndt97} was to lubricate the inside
of the cavity, thus hindering the H bonds responsible for the
artificial slowing down; our strategy was the make the surrounding
system softer. In the experimental case the effect was clear:
lubricated cavities are faster than unlubricated ones; smaller
cavities are faster than larger cavities. In our case, we also obtain
that FS cavities are faster than FC cavities; whether or not smaller
cavities are faster than larger ones is unclear, but the data do not
rule this out.

The differences are also relevant. In the experimental case the
confined (free) system is liquid salol, and the pore is glass. Hence,
even though one may say that there are amorphous boundary conditions,
these are certainly not drawn from the Gibbs-Boltzmann equilibrium
distribution of an external salol system. Moreover, the reasons for
the original `stiffness' are also different. In the experimental case
it is the formation of H bonds between internal salol and the surface
of the pore. In our case, the nature of the bonds between particles
within the cavity and across the interface is exactly the same;
however, the complete freezing of the cavity suppresses the swap,
uncooperative, rearrangements useful to reach
equilibrium. Accordingly, the solutions adopted are also different.

We cannot not say whether or not the similarities overcome the
differences, so to make this experimental case significant to our
context. We limit ourselves to register the fact that the problem of
an artificial slowing down in confining geometries has already
occurred in experiments and that, when solved, the cavity dynamics can
change very dramatically.

%%%%%%%%%%%%%%%%%%%%%%%%%%%%%%%%%%%%%%%%%%
\section{Conclusions}
%%%%%%%%%%%%%%%%%%%%%%%%%%%%%%%%%%%%%%%%%%

\label{sec:conclusions}

We have studied the dynamics of a confined cavity, using different
Montecarlo algorithms and different amorphous boundary conditions. Our
bare findings are:
\begin{enumerate}
\item FC---swap---low $T$: the cavity relaxation time $\tau$ is larger
  the larger $R$ and it saturates at $R\sim \xi_s$, where $\xi_s$ is
  the point-to-set correlation length.
\item FC---swap: in the region $R\sim \xi_s$ a colder cavity relaxes
  faster than a hotter cavity.
\item FC---swap: at higher $T$ the relaxation time $\tau(R)$ displays
  an overshooting that disappears on lowering $T$.
\item FC---nonswap: $\tau$ is larger the smaller $R$.
\item FC---nonswap: small cavities ($R< 4$) are completely stuck at an
  off-equilibrium level.
\item FS nonswap dynamics is significantly faster than FC nonswap
  dynamics; with FS small cavities are no longer stuck.
\item The FS point-to-set correlation function $q(R)$ is nonzero in
  the region of interest of $T$ and $R$.
\item FS---nonswap: data are compatible with a nonmonotonic $\tau(R)$;
  data do not rule out the possibility that in small cavities $\tau$
  is smaller the smaller $R$.
\end{enumerate}

We have proposed a theoretical scenario whose aim is to organize
all these results into one coherent picture. Our scenario rests on two
main ideas. First, depending on the values of $R$ and $T$, and on the
type of dynamics, there may be an hybridization between MCT and
activated relaxation channels; this hybridization, when present,
gives rise to a nonmonotonic cavity relaxation time $\tau(R)$. Second,
the frozen configuration setup is unsuitable to run nonswap dynamics,
and in general it is not very physical, as it may give rise to an
artificial dynamical freezing. We have introduced a frozen state setup,
based on the idea that the amorphous boundary condition must select a
certain state, not simply a certain configuration. If we trust these
two ideas, then we can find an interpretation for the very diverse
results we find.

Result 1 supports the concept that $\xi_s$ is the relevant scale of
cooperativity in the system. According to the RFOT
theory with fluctuating surface tension, the activated relaxation time
is equal to the in-state relaxation time for $R\ll\xi_s$, it grows
when $R$ gets across the support of the probability distribution of
the rearranging sizes $P(R,\xi_s)$,  and
it finally saturates to its bulk value for $R\gg\xi_s$. Hence, when
the cavity is larger than the scale of cooperativity relaxation
factorizes, whereas when the cavity is smaller than $\xi_s$ the whole cavity must
rearrange collectively. This RFOT interpretation is supported by
result 2: an inversion of the relaxation time (cooler is faster) 
happens because a colder cavity may still be confined within just one
state, thus showing only the short, in-state relaxation time, while
(at the same value of $R$) a hotter cavity may be already unlocked,
thus sporting the full bulk relaxation time.

The maximum displayed by the local swap $\tau(R)$ (result 3) is one
piece of evidence in support of the (rather speculative) scenario described in
Section VIII: the hybridization between nonactivated MCT channels and
activated channels gives rise in the bulk to the crossover between MCT
and activation close to $T_c$, while in the cavity it gives rise to a
nonmonotonic $\tau(R)$. This hybridization implies that for large $R$
the cavity relaxation time follows the MCT branch, so that $\tau$ is
larger the smaller $R$, which is in agreement with the nonswap
dynamics result 4. On the other hand, switching to swap dynamics has a
twofold effect: in the bulk, it eliminates the $T_c$ crossover; in the
cavity, it flattens the maximum of $\tau(R)$.

We have speculated that the complete freezing out of small cavities
with nonswap dynamics (result 5) is not quite physical, and we have
suggested that it could be the effect of an artificial suppression of
some elastic (noncooperative) relaxation modes due to the frozen
configuration setup. We have proposed a practical way to implement
amorphous boundary conditions with a frozen state and we have found
that this setup speeds up significantly the nonswap dynamics,
unlocking the small cavities (result 6). We have also checked that the
point-to set correlation remains nonzero, despite a significant
reduction due to the smaller degree of confinement by the external
state (result 7).

Finally, we tried to understand what was the behavior of $\tau$ as a
function of $R$ in the frozen state case. This is quite crucial: if we
cannot find {\it any} regime of $R$ and $T$ where the nonswap $\tau$
is smaller for smaller $R$, then we have a problem. Our entire
construction relies on the idea that for small enough $R$ the MCT
branch must be gone, so that all that remains is the activated branch,
and this must be faster the smaller the cavity. Our time series
(Fig.~\ref{fig:punch}) are too short to settle this matter. But we can
at least say that the data do not rule out this possibility (result
8). With a little more optimism, we can even conclude that the
smallest cavity is not the slowest one, which is all we need to
support our theoretical scenario.

The whole scenario still admits considerable improvements in
clarity. As we have said, longer simulation with nonswap dynamics in
the FS setup are needed to study carefully $\tau(R)$, and this should
be done at several values of $R$ and of $T$. At the same time, FS swap
simulations should be run in order to reconstruct the entire
point-to-set correlation function, $q(R)$, to check whether or not it
retains its essential properties. Is it still a nonexponential
function at lower temperature?  How does the FS correlation length
$\xi_s$ compare to its FC counterpart?  Work in this direction is in
progress.

 \acknowledgments We thank
L.~Berthier, G.~Biroli, J.-P.~Bouchaud, C.~Cammarota, L.~Cugliandolo,
S.~Franz, J.P.~Garrahan, I.~Giardina, G.~Gradenigo, R.L.~Jack, A.~Heuer, W.~Kob,
M.~Mezard, G.~Parisi, G.~Tarjus, M.~Wyart and F.~Zamponi for several
important remarks, and ECT* and CINECA for computer time. The work of
TSG was supported in part by grants from ANPCyT, CONICET, and UNLP
(Argentina). PV has been partly supported through Research Contract
Nos. FIS2009-12648-C03-01,FIS2008-01323 (MICINN, Spain).

%% Bibliography %%%%%%%%%%%%%%%%%%%%%%%%%%%%%%%%%%%%%%%%%

\bibliographystyle{apsrev}
\bibliography{all}

\end{document}